\documentclass[a4paper,12pt]{article}
\usepackage[top=1.25in, bottom=1.25in, left=1.05in, right=1.05in]{geometry}
\usepackage[utf8]{inputenc}
\usepackage[T1]{fontenc}
\usepackage[english]{babel}

\usepackage{natbib,longtable}
\usepackage[flushleft]{threeparttable}
\usepackage{threeparttablex} 
\usepackage{eurosym}

\usepackage[section]{placeins}

\usepackage{setspace}

\usepackage{enumitem}
\setlist{noitemsep}  

\usepackage[colorinlistoftodos,color=orange!60]{todonotes}
\usepackage{regexpatch}
\makeatletter
\xpatchcmd{\@todo}{\setkeys{todonotes}{#1}}{\setkeys{todonotes}{inline,#1}}{}{}
\makeatother

\usepackage{amscd,amsmath,amssymb,amsfonts,amsthm,bbm,bm,latexsym,mathrsfs,mathtools,bigints}
\usepackage[subnum]{cases}	
\usepackage{dsfont}
\makeatletter
\newcommand*{\distas}[1]{\mathbin{\overset{#1}{\kern\z@\sim}}}	
\makeatother
\newcommand*\abs[1]{\left|#1\right|}		
\allowdisplaybreaks

\newtheorem{theorem}{Theorem}

\theoremstyle{plain}
\newtheorem{remark}{Remark}

\RequirePackage[colorlinks=true, citecolor=blue, urlcolor=blue]{hyperref}

\usepackage{xcolor,subfigure,epsfig,epstopdf,rotating,graphics,graphicx} 
\graphicspath{{./figures/}}
\usepackage[width=0.9\textwidth, font={footnotesize}]{caption}

\usepackage{rotating,lscape,pdflscape}
\usepackage{booktabs,longtable,float,array,multirow,colortbl,adjustbox}

\newcolumntype{C}[1]{>{\centering\arraybackslash}p{#1}}

\makeatother

\makeatletter

\def \R {\mathds{R}}

\DeclareMathOperator*{\argmin}{argmin}

\makeatother

\title{\vspace{-60pt} \textbf{Comparing predictive ability in presence of instability over a very short time}
\thanks{\footnotesize
The authors gratefully acknowledge the participants at the 31st SNDE Symposium in Padova, Fondazione Eni Enrico Mattei, and the 2024 European Association of Young Economists (EAYE) annual meeting in Paris for their helpful feedback. The authors also gratefully acknowledge Laura Coroneo for her useful comments and feedback. This research used the Computational resources provided by the Core Facility INDACO, which is a project of High-Performance Computing at the University of Milan. Fabrizio Iacone and Luca Rossini acknowledge financial support from the Italian Ministry of University and Research (MUR) under the Department of Excellence 2023-2027 grant agreement ``Centre of Excellence in Economics and Data Science'' (CEEDS).
}
}

\author{
Fabrizio Iacone\thanks{Universit\`{a} degli Studi di Milano and University of York, \color{blue}\texttt{fabrizio.iacone@unimi.it}}
\and
Luca Rossini\thanks{Universit\`{a} degli Studi di Milano and Fondazione Eni Enrico Mattei (FEEM), \color{blue}\texttt{luca.rossini@unimi.it}}
\and 
Andrea Viselli\thanks{Universit\`{a} degli Studi di Milano,
 \color{blue}\texttt{andrea.viselli@unimi.it}}
}

\date{\today}

\begin{document}
\maketitle

\begin{abstract}

We consider forecast comparison in the presence of instability when this affects only a short period of time. 
We demonstrate that global tests 
do not perform well in this case, as they were not designed to capture very short-lived instabilities, and their power vanishes altogether when the magnitude of the shock is very large. 
We then discuss and propose approaches that are more suitable to detect such situations, such as nonparametric methods (S test or MAX procedure). 
We illustrate these results in different Monte Carlo exercises and in evaluating the nowcast of the quarterly US nominal GDP from the Survey of Professional Forecasters (SPF) against a naive benchmark of no growth, over the period that includes the GDP instability brought by the Covid-19 crisis.
We recommend that the forecaster should not pool the sample, but exclude the short periods of high local instability from the evaluation exercise. 

\end{abstract}

\clearpage

\doublespacing

\section{Introduction}
    \label{intro}

Instabilities during periods of crisis are common in time series and in particular in the last years when the COVID-19 pandemic caused a huge shock and changed the economic perspective. Consequently, forecasting becomes more challenging during periods of crisis and sudden recoveries, but these periods are often more important in the forecasting task since they usually carry more risk of catastrophic errors.

An exceptional example is the methodology proposed by \citet{bok2018macroeconomic}, which was implemented by the New York FED but was suspended in September 2021 due to the challenges posed by the COVID-19 pandemic and recently was reintroduced as in \cite{almuzara2023new}. In particular, the period related to COVID-19 and the sudden recovery \citep[see, e.g.][]{Ng2021, Lenza2022} has gained attention among researchers and policymakers since for the first time we have had the opportunity to study some unpredictable shocks in the economy.

In detail, COVID-19 caused a shock to GDP that was in many ways unprecedented in recent history and tested the capacity of current methodologies in the presence of extraordinary situations. The main big challenges are both in how the forecasting methods behave but also how we should evaluate the forecasting that common models provide the policymakers. The first challenge has been strongly studied in the literature, where \cite{huber2023nowcasting} argue that nonlinear methods may better accommodate extreme situations. Differently, \cite{foroni2022forecasting} and \cite{Schor2021} stress the resilience of popular models, such as mixed frequency or dynamic factor models to accommodate the severity of recessions. 

Indeed, the second task has not been strongly discussed in the literature, but it reveals the importance of evaluating and accounting forecasting instabilities. In such situations, average tests for forecast evaluation, such as the Diebold and Mariano test of equal unconditional predictive ability \citep[see,][]{dieboldComparingPredictiveAccuracy2002, GiacominiWhite2006} may not be informative, as they do not have much power to detect instances in which one forecast outperforms the competitor only on a fraction of the sample. 
To account for forecasting instability, \citet{rossi2021forecasting} recommends the application of local procedures, like the one-time reversal or the fluctuation tests from \citet{giacomini2010forecast}. 
In comparison with diagnostics based on the evaluation of forecasts over an average, the fluctuation test is indeed local, as the statistic is only computed over a fraction of the out-of-sample evaluation period.

However, as the test is derived assuming that the fluctuation induced by the instability spans a relevant fraction of the sample, then it replicates on a smaller scale the difficulties incurred by the average, global \cite{dieboldComparingPredictiveAccuracy2002} test. Indeed, the Monte Carlo study in \citet{giacomini2010forecast} shows that the attempt to run the test for a very short fluctuation period is frustrated by a relevant size distortion. 


In this paper, we, therefore, discuss the difficulties associated with using global diagnostics for evaluation in times of crises, and we consider other diagnostics, based on the predictive instability tests in \citet{andrews2003end} and \citet{Taylor_etal_2021_Pockets}, to account for this situation. Indeed, situations of crises like the one induced by COVID-19 or by an economic recession do not typically conform well with the assumption of a large, even if local, evaluation period, as crises often span only a very small number of observations. As an example, in the COVID-19-induced recession, the instability mostly affected just two or three quarters. 

One key finding of our paper is that in the presence of high, very localised instability, global tests may have no power at all, thus leading to the incorrect conclusion, and that this may also obfuscate a signal that would otherwise be clear, had the analysis not included the brief period of instability. In particular, we investigate in different Monte Carlo exercises the importance of considering the S test and the MAX procedure instead of the usual global tests when we have a short deviation. Moreover, we emphasize the importance of applying a correction for the dependence in the S test statistic by using the estimated value using the pre-changed sample and not the recommended in \cite{andrews2003end} or a simple identity matrix.

In an empirical application to the US nominal GDP growth, we compare the nowcast from the Survey of Professional Forecasters (SPF) and the last available observation during the COVID-19 recession and the subsequent recovery. When COVID-19 is not considered in the analysis, DM and Fluctuation tests suggest that the SPF nowcast is more precise than the naive model, which does not happen when we include the COVID-19 period. However, if we consider the $S$ test and the MAX procedure, it provides more appropriate results.
Given these results, we recommend that the forecaster should not pool the sample, but exclude the short periods of high local instability from the evaluation exercise.

The rest of the paper is organised as follows. In Section~\ref{Sec_statistics}, we introduce the Diebold and Mariano (DM) and Fluctuation test statistics, and we investigate their performance in case of a very large, localised shock. We also introduce some diagnostics to detect the presence of those shocks.  
In Section~\ref{Sec_MC} we study the performance of these diagnostics in two different Monte Carlo exercises; while in Section~\ref{Sec_Application} we evaluate the ability of the Survey of Professional Forecasters (SPF) to outperform a naive benchmark when a period of high but localised instability due the Covid-19 shock is included in the study. Section~\ref{Sec_Conclusions} concludes the paper.

\section{Detecting forecast breakdowns over very small samples}\label{Sec_statistics} 

\subsection{Description of the environment}\label{environment}

We consider the classical \citet{GiacominiWhite2006} framework, also see for example \citet{giacomini2010forecast}, or \citet{Coroneo2020Comparing}. To fix some notation, we denote the variable of interest by $y_t$, for which we want to compare two $h$-step ahead forecasts obtained from two alternative forecasting methods, based on some predictor variables $x_{t}$.
We denote the observed vector by $w_t \equiv (y_t, x'_t)'$, defined on a complete probability space $(\Omega,\mathcal{F}, P)$, and  the information set at time $t$ by $\mathcal{F}_t=\sigma(w_1',\hdots, w'_t)$.
The two $h$-step ahead forecasts for time $t$ are based on the information set $\mathcal{F}_{t-h}$ and are denoted by $\widehat{y}^{(i)}_{t}\left(\widehat{\delta}^{(i)}_{t-h, R_i}\right)\equiv f^{(i)} \left(w_{t-h}, w_{t-h-1}, \hdots, w_{t-h-R_i+1};\widehat{\delta}^{(i)}_{t-h, R_i}\right)$  for $i=1,2$, where the forecasts are measurable functions of a sample of size $R_1$ for $f^{(1)}$ and $R_2$ for $f^{(2)}$. If a forecast is based on parametric models, the vector $\widehat{\delta}^{(i)}_{t-h, R_i}$ includes the estimates from the model. Otherwise, $\widehat{\delta}^{(i)}_{t-h, R_i}$ represents the semiparametric or nonparametric estimator used to construct the forecast. Notice, however, that in this framework the estimates  $\widehat{\delta}^{(i)}_{t-h, R_i}$ are based on a rolling window of dimension $R_i<\infty$, so they are inconsistent.  

For the two forecasts $\widehat{y}^{(i)}_{t}\left(\widehat{\delta}^{(i)}_{t-h,R_i}\right)$, denote the forecast error by 
$e^{(i)}_{t}\left(\widehat{\delta}^{(i)}_{t-h, R_i}\right)= y_t - \widehat{y}^{(i)}_{t}\left(\widehat{\delta}^{(i)}_{t-h, R_i}\right)$ and, for a real function $L(\cdot)$, that we interpret as a loss function, the loss associated with the forecast error is $L\left(e^{(i)}_{t}\left(\widehat{\delta}^{(i)}_{t-h, R_i}\right)\right)$. Finally, the loss differential at time $t$ between the two forecasts is 
\begin{equation*}
d_{t}\left(\widehat{\delta}^{(1)}_{t-h,R_1}, \widehat{\delta}^{(2)}_{t-h,R_2}\right)=
L\left(e^{(1)}_{t}\left(\widehat{\delta}^{(1)}_{t-h,R_1}\right)\right)- L\left(e^{(2)}_{t}\left(\widehat{\delta}^{(2)}_{t-h,R_2}\right)\right),
\end{equation*} 
and the null hypothesis of equal predictive ability of the two forecasting methods is
\begin{equation}\label{EPA_null}
H_0 : E \left( d_{t}\left(\widehat{\delta}^{(1)}_{t-h,R_1}, \widehat{\delta}^{(2)}_{t-h,R_2}\right) \right) =0
\end{equation}
at each point $t$. Denoting $R \equiv \max(R_1,R_2)$, we assume that we have a sample of dimension $R+h+T-1$ and we can therefore evaluate the hypothesis \eqref{EPA_null}  in the $\{R+h,\ldots,R+h+T-1 \}$ evaluation period. To abbreviate notation, denote $s=t-(R+h)+1$ and 
\begin{equation*}
d_{s}\equiv d_{t}\left(\widehat{\delta}^{(1)}_{t-h,R_1}, \widehat{\delta}^{(2)}_{t-h,R_2}\right),
\end{equation*} 
where notice that the evaluation period with respect to $s$ is $\{1, \ldots, T\}$.
\\
Let us denote $\mu_s$ by 
\begin{equation*}
\mu_s=E(d_s).
\end{equation*}
Then, the null hypothesis of equal predictive ability of the two forecasting methods is
\begin{equation}\label{h0}
H_0 : \mu_s =0
\end{equation}
at each point $s \in \{1, \ldots, T\}$.\\ When $H_0$ is not met, then there is a $\mu_s \neq 0$ for some $s$ in the evaluation period.
 
\subsection{Diebold and Mariano equal predictive ability test}\label{sub_DM}

\citet{dieboldComparingPredictiveAccuracy2002} propose to test the hypothesis in \eqref{h0} using the sample average 
\begin{equation*}
\overline{d}=\frac{1}{T}\sum _{s=1}^{T}d_{s}.
\end{equation*}
Denoting the long-run variance by
\begin{equation*}
\sigma^2_T=Var \left( \frac{\sqrt{T}}{T}\sum _{s=1}^{T}d_{s} \right)
\end{equation*}
and by $\widehat{\sigma}_T^2$ an estimate of $\sigma^2_T$, the Diebold and Mariano test (hereafter DM test) uses the statistic
\begin{equation*}
t_{DM}=\sqrt{T}\frac{\overline{d}}{\widehat{\sigma}_T}. 
\end{equation*} 
When $\widehat{\sigma}^2_T-\sigma^2_T=o_p(1)$ and given other regularity conditions (see for example Assumption GW in Subsection~\ref{sec:GWtest}), \citet{GiacominiWhite2006} show that, under $H_0$, 
\begin{equation*}
t_{DM} \rightarrow_d Z, 
\end{equation*} 
where $Z$ is a standard normal distributed variable. {When the null hypothesis is not met, the DM test has non-trivial power in presence of local alternatives $\mu_s=\delta T^{-1/2}$ for all $s$.}

\subsection{Giacomini and Rossi fluctuation test}\label{sub_fluctuation}
\citet{giacomini2010forecast} describe the DM test as an average or global test, as it primarily detects deviations of the null hypothesis that are constant over the whole evaluation period. The DM test is less effective in detecting deviations from $H_0$ when they occur only on a fraction of the sample, and this loss of local power is stronger the shorter the fraction where the predictive ability of the two forecasts is different. Hence, the DM test might even have no power at all when $\mu_s$ changes sign over the evaluation sample so that $\sum_s \mu_s =0$ is possible. 

Therefore, \citet{giacomini2010forecast} propose to consider a local statistic, called the fluctuation statistic 
\begin{equation*}
Fl_{s,k}=\frac{\sqrt{k}}{k}  \frac{1}{\widehat{\sigma}}\sum_{l=s-k/2}^{s+k/2-1}d_l,
\end{equation*} 
where $k=\lfloor \kappa T \rfloor$ and we assume $k/T \to \kappa \in (0,\infty)$ as $k \to \infty$ and $T \to \infty$ as in Assumption 1(c) in \cite{giacomini2010forecast}. They show that, under $H_0$ and regularity conditions,
\begin{equation*}
Fl_{s,k} \Rightarrow \frac{B(\rho+\kappa/2)-B(\rho-\kappa/2)}{\sqrt{\kappa}},
\end{equation*}
where $B(\cdot)$ is a standard univariate Brownian motion and  $\rho \in [0,1]$ is such that $s=\lfloor \rho T \rfloor$.
The fluctuation test statistic is defined as
\begin{equation*}
FL_\kappa=max_s \abs{Fl_{s,k}}, 
\end{equation*} 
hence \citet{giacomini2010forecast} characterise the convergence to the limit of the test statistic and provide simulated critical values for the test.
In comparison with the DM test, the fluctuation test should have less power when $\mu_s$ is constant, but more when the predictive ability is different only on a subsample.

It is noteworthy that the critical values for the fluctuation test depend on the length of the fraction $\kappa$: the Monte Carlo study in \citet{giacomini2010forecast} suggests a certain size-power trade-off in the choice of $\kappa$, in the sense that very small values ($\kappa=0.1$) are associated to size distortion in finite samples, whereas larger values are associated to lower power in presence of instability.

\subsection{Equal predictive ability tests in presence of brief events} \label{sec:GWtest}
 \citet{giacomini2010forecast} demonstrate the value of the fluctuation test in the forecasting of exchange rate macroeconomic models. As their example makes clear, the natural application is in situations where the economic dynamics are slowly changing over time, and we can use the test to study the evolution, as in \citet{Rossi2010Application} and \citet{Galvao2021JudgementalAdjustment}. In other words, the test detects forecast differential instability across periods, or regimes, as it happens when forecasts from a macroeconomic model are compared against a benchmark in the presence of changes to the fundamental economic relations. 
 
 The fluctuation test is also effective in detecting the existence of what \citet{Timmermann2008Elusive}, emphasising the local nature of the predictability of returns,  refers to as pockets of predictability, that only appear in some periods in time corresponding to fractions in the sample, as in  \citet{Urga2023_Pockets_Application}.
This situation, however, does not cover well the differential forecasting instability that occurs over only a small period of time, as is sometimes the case for economic recessions or other short-lived events. 

In this case, the expected value of the differential predictive ability is rather described as 
\begin{equation*}
\mu_s=\delta_1 T^{-1/2}+\delta_2 T^a I_s(\tau),    
\end{equation*}
where $I_s(\tau)$ is an indicator function, taking value $1$ if $s=\lfloor \tau T \rfloor$ and $0$ otherwise; the factor $\delta_2 T^a$ characterises the dimension of the change in the prediction differential in relation to the sample size. 

 We then assume for the loss differential $d_s$ the data generating process
\begin{equation}\label{regression}
    d_s=\delta_1 T^{-1/2}+\delta_2  T^a I_s(\tau)+u_s,
\end{equation}
  where $u_s$ is a zero-mean process. In this case, the loss differential $d_s$ is constant, but for a point in time $s=\lfloor \tau T \rfloor$ the situation of equal predictive ability corresponds to $\delta_1=0$ and $\delta_2=0$. 

  We first study the limit properties of the DM statistic assuming that the long-run variance is estimated using the Bartlett kernel,
  \begin{equation*}
    \widehat{\sigma}_T^2=c_0+2 \sum_{l=1}^M {\frac{M-l}{M} c_l},  
  \end{equation*}
  where $c_l$ is the $l$-th sample covariance of $d_s$ (with $l=0$ the sample variance) and $M$ is a user-chosen bandwidth such that $M/T \rightarrow 0$ as $T \rightarrow \infty$.
To establish the limit properties of the DM statistic, we introduce the following assumptions

\textbf{Assumption GW} \textit{
\begin{itemize}
    \item[(GW.1)] $u_s$ is mixing with $\phi$ of size $-r/(2r-2)$, $r \geq 2$; or $\alpha$ of size $-r/(r-2)$, $r>2$; 
    \item[(GW.2)] $E\left(\left| u_s \right| ^{2r}\right)<\infty$ for all $s$;
    \item[(GW.3)] $Var \left( \frac{\sqrt{T}}{T}\sum _{s=1}^{T} u_{s} \right)>0$ for all $T$ sufficiently large.
\end{itemize}
}

\begin{remark}\label{rem_GW1}
When $\delta_2=0$, Assumption $GW.1$ may be formulated in terms of $w_s$, and $GW.2$ and $GW.3$ in terms of $d_s$, as in  \citet{GiacominiWhite2006}, to which we refer for a discussion of these assumptions.  
\end{remark}
To characterise the local power of the test, it is also convenient to assume that there is $\sigma^2 = \lim_{T\rightarrow \infty}Var \left( \frac{\sqrt{T}}{T}\sum _{s=1}^{T} u_{s} \right)$.
\begin{theorem}\label{DM} 
Under Assumptions $GW.1$ -- $GW.3$, 
\begin{itemize}
    \item[(i)] if $a<1/2$, then $t_{DM} \rightarrow_d Z+\frac{\delta_1}{\sigma}$;
    \item[(ii)] if $a>1/2$, then $|t_{DM}| \rightarrow_p 1$.
\end{itemize}
  \end{theorem}
  We refer to Appendix~\ref{sec:AppA} for a complete proof.
\begin{remark}\label{rem_Thm_dm}
The DM test is therefore not able to detect superior forecasting ability when these are limited to just one point in time. In fact, in case of very large differentials, the power of the test drops to 0. \\
As the Fluctuation test uses a fraction of the sample size that is proportional to the whole sample, and the same estimate for the long-run variance, qualitatively similar results also hold for the Fluctuation test. 
\end{remark}
A similar argument of course holds when the instability affects more than one observation, provided that the number is very small relative to the sample. The COVID-19-induced instability seems the typical example of this situation, but it suggests that including in the evaluation exercise the recession induced by the financial crisis may also generate a power loss, although this should be more subdued as the size of the shock is less. 

\subsection{Detecting predictive superiority in presence of brief, extreme instability}

Detecting forecasting superiority is therefore more difficult in cases of events that are limited in time and large in scale, and yet it is also usually more important for its policy implications.
When the location of the event is known in advance, as it may be in the case of the recession induced by COVID-19, we propose to apply the predictive instability test of \citet{andrews2003end}, and we show that its application to evaluate forecasts is justified. When the location of the potential predictive instability is not known, the approach is rather similar to detecting outliers or extreme values, as in \cite{Leadbetter1983}.

\subsubsection{Predictive instability test when the location is known, the $S$ test}

The test defined in \citet{andrews2003end} is based on testing the null hypothesis $H_0: \delta=0$ in 
\begin{equation}\label{unrestr}
    d_s=\mu +\delta  I_s(\tau)+u_s, 
\end{equation}
where $I_s(\tau)$ is treated as a dummy variable taking a non-zero value only when $s=\lfloor \tau T\rfloor$, and $u_s$ is a zero-mean process. \\
To simplify notation, we assume $\tau=1$, as in \citet{andrews2003end}, where the generic $\tau$ situation is also briefly discussed.  

In general, comparing residuals sum of squares from an unrestricted and restricted regression is done using an $F$ test or, if the data are not normally distributed, using a $\chi^2$ limit distribution. However, in this case, the usual asymptotic convergence in distribution of the $F$ statistics to a $\chi_1^2$ limit does not hold: intuitively, this is because $\delta$ is estimated using only one observation, so it is not possible to invoke a central limit theorem to establish the limit distribution of this estimate.

Instead, denoting $\widetilde{\mu}=\frac{1}{T} \sum_{s=1}^T d_s$, the estimate of $\mu$ in the restricted model, and $\widehat{\mu}=\frac{1}{T-1}\sum_{s=1}^{T-1} d_s$ the estimate in the unrestricted model, so that the restricted and unrestricted residuals are $\widetilde{u}_s=d_s-\widetilde{\mu}$ and $\widehat{u}_s=d_s-\widehat{\mu}$, respectively. Then the idea is to estimate the distribution of $\widetilde{u}_{T}^2$ with the sample distribution of $\widehat{u}_s^2$ for $s=1,\ldots, T-1$.   

To improve the empirical size performance in finite sample, \citeauthor{andrews2003end} proposes a slight modification of this procedure, where the critical distribution is estimated from $\widehat{u}_{2(s)}=d_s-\widehat{\mu}_{2(s)}$, where $\widehat{\mu}_{2(s)}=\frac{1}{T-2} \sum_{j=1, j\neq s}^{T-1} y_j$.
The null hypothesis is rejected at $\alpha$ asymptotic significance level if  $\widetilde{u}_{T}^2$ exceeds the $(1-\alpha)$ sample quantile of 
$\widehat{u}_{2(s)}^2$ for $s=1,\ldots, T-1$. 

To state the properties of the test, we introduce some additional notation, adapting the one presented in \citet{andrews2003end}. 
Denote the test statistic by $S$, so $S=\widetilde{u}_{T}^2$, and, for any $s \neq T$, let $S_s(\mu)=d_s-\mu=u_s$: under strict stationarity, all these variables have the same distribution, denoted as $F_S(x)$. Also, let $S_s=\widehat{u}_{2(s)}^2$, with empirical distribution 
$\widehat{F}_{S,T}(x)=\frac{1}{T-1}\sum_{s=1}^{T-1} 1(S_s \leq x)$, and let $q_{S,1-\alpha}$ denote the $(1-\alpha)$ quantile of $F_S(x)$, and $\widehat{q}_{S,1-\alpha}$ denote the $(1-\alpha)$ sample quantile of  $S_{s,s\neq T}$.  
Finally, let $S_{\infty}$ be a random variable with the same distribution as $d_{T}-\mu$. 

Then we introduce the following assumptions:

\textbf{Assumption A} \textit{
\begin{itemize}
    \item[A.1] $w_t$ is strictly stationary for all $t$ if $H_0$ holds, and for $s\neq T$ otherwise;
    \item[A.2] $w_t$ is ergodic for all $t$ if $H_0$ holds, and for $s\neq T$ otherwise;
    \item[A.3] $E(u_1)^2<\infty$;
    \item[A.4] $u_1$ has absolute continuous distribution at the quantile $\alpha$.    
\end{itemize}
}
In Assumptions A.3 and A.4, we refer to $u_1$ as a generic $u_s$ as the strict stationarity of $w_t$ and the nature of the forecasting functions ensure the strict stationarity of $u_s$.

\begin{theorem}\label{Thm_Andrews}
Under assumptions A.1-A.4, as $T \rightarrow \infty$,
\begin{itemize}
    \item[(i)] $S \rightarrow_d S_\infty$ as $T \rightarrow \infty$ under $H_0$ and $H_1$;
    \item[(ii)] $\widehat{F}_{S,T}(x) \rightarrow_p F_S(x)$ under $H_0$ and $H_1$;
    \item[(iii)] $\widehat{q}_{S,1-\alpha} \rightarrow_p q_{S,1-\alpha}$ under $H_0$ and $H_1$;
    \item[(iv)] $P \left(S > \widehat{q}_{S,1-\alpha} \right) \rightarrow \alpha$ under $H_0$.
\end{itemize} 
\end{theorem}
  We refer to Appendix~\ref{sec:AppA} for a complete proof.

\begin{remark}\label{rem_Andrews1}
Assumptions $A.1$ and $A.2$ are in terms of the observables $w_t=\left(y_t,x_t'\right)'$. As in \cite{andrews2003end}, they are only referred to the period of stability. 
\\
The restriction to stationarity rules out heterogeneity in the distribution, including heteroskedasticity, and in this sense it is stronger than requirements in \citet{GiacominiWhite2006} or in Theorem~\ref{DM}, where mixing is allowed. That was possible since the limit distribution of the test statistics was derived using central limit theorem arguments. In this case, we use the assumption of identical distribution to estimate the distribution of the residuals. \\
Assumptions $A.3$ and $A.4$ are in terms of $u_s$: this is equivalent to (b) and (d) in Assumption LS in \cite{andrews2003end}: as similar assumptions on the unobservable term in \citet{GiacominiWhite2006} and \citet{giacomini2010forecast}, these can be investigated on a case by case basis.
\end{remark}
\begin{remark}\label{rem_Andrews3}
The presentation and the statement of Theorem~\ref{Thm_Andrews} allow for instability only at the end of the sample; this follows the outline in \cite{andrews2003end}. Instabilities at different points in time can also be considered, as it is discussed in  \cite{andrews2003end}.  
\end{remark}

In the interest of simplicity, we present Theorem~\ref{Thm_Andrews} assuming that the instability only affects one point. Of course, it is possible to consider a longer span of observations, as long as this remains finite and, in practice, also small concerning the sample size. When instability affects $k$ observations ($T-k,\ldots,T$),  \cite{andrews2003end} shows that the procedure can be easily applied to quadratic forms of $\widetilde{U}_s=(\widetilde{u}_{s},\ldots,\widetilde{u}_{s+k-1})'$, $\widehat{U}_{2(s)}=(\widehat{u}_{2(s)},\ldots,\widehat{u}_{2(s+k-1)})'$, adjusting the definition of $\mu_{2(s)}$ to account for the fact more residuals are considered jointly.  

One could compute quadratic forms directly from $\iota_k'\widetilde{U}_s$ and $\iota_k'\widehat{U}_{2(s)}$, where $\iota_k$ is a $k\times1$ vector of ones.  \cite{andrews2003end} also proposes an alternative statistic, that would account for the autocorrelation in $u_s$: denoting $\Sigma$ the variance-covariance matrix of ${U}_s=({u}_{s},\ldots,{u}_{s+k-1})'$, this alternative quadratic form is computed from $\iota_k'\Sigma^{-1}\widetilde{U}_s$ and $\iota_k'\Sigma^{-1}\widehat{U}_{2(s)}$.

\begin{remark}\label{rem_Andrews5}
 As $\Sigma^{-1}$ is unobservable, \cite{andrews2003end} proposes to estimate $\Sigma$ using the restricted residuals $\widetilde{u}_s$ over the whole sample.  We denote this estimate as $\widetilde{\Sigma}$. \\
\cite{andrews2003end} assumes that the explanatory variables do not depend on the sample size ($T$), ruling out in the regression $\eqref{unrestr}$ a factor proportional to $T^\alpha$, as in $\eqref{regression}$. Thus, the estimates $\widetilde{\mu}$ and $\widetilde{\Sigma}$ are consistent, see Lemma 1 in \cite{andrews2003end}. When however model \eqref{regression} is correct, the estimate $\widetilde{\Sigma}$ may fail to be consistent. This is clearly a relevant issue in our situation, and we explore it further in the Monte Carlo experiment. In this case, a consistent estimate of $\Sigma$ may still be obtained using the residuals from the regression in the stability part of the sample only, $\widehat{u}_s$, and we denote it as $\widehat{\Sigma}$.   
\end{remark}

\subsubsection{Predictive instability when the exact location is not known, the \textit{MAX} diagnostic}

The test in \cite{andrews2003end} is designed for situations in which the exact location of the suspected instability is known. In many cases, however, the exact location of the occurrence of a brief differential in predictive is not known in advance. This situation is rather more similar to the problem of detecting one of the "pockets of predictability" discussed in \citet{Timmermann2008Elusive}, as the forecast of a model is measured against a benchmark of no predictive ability. 

A recent procedure to detect such pockets is in \citet{Taylor_etal_2021_Pockets}, where it is characterised as the occurrence of an outlier at an unknown point in time in a series. As they are interested in detecting the location of a pocket, as well as in testing its statistical significance, \citet{Taylor_etal_2021_Pockets} divide their sample into many short intervals and compute a $t$ statistic for each one. Our problem of interest is somewhat simpler, as we can just look at the statistic $d_s^2$.   

The procedure consists of splitting the sample in a training period, $s=1,\ldots, T^*$, and a monitoring period $s=T^*+1,\ldots, E$ where $E \leq T$, and $T^*$ and $E$ are fractions of the sample period $T^*=\lfloor \lambda_1 T \rfloor$, $E=\lfloor \lambda_2 T \rfloor$, for $0 < \lambda_1 < \lambda_2 \leq 1$. 
We assume that no instability occurs during the training period, but instability may take place during the monitoring period. 
We then compare the maximum of the statistic $d_s^2$ during the training period, $\max_{s=1,\ldots,T^*} d_s^2$,  and during the monitoring period, $\max_{s=T^*+1,\ldots,E} d_s^2$. In practice, we use the training period to estimate the probability that $\max_{s=T^*+1,\ldots, E} d_s^2> \max_{s=1,\ldots, T^*} d_s^2 $ if no instability has occurred.  

Notice that, heuristically, if the statistic of interest $d_s^2$ was independently and identically distributed, and the training and monitoring period constituted fractions $\lambda_1$ and $\left( 1-\lambda_1 \right)$ of the sample, respectively, then in large sample that probability should be $\left( 1-\lambda_1 \right)$. \citet{Taylor_etal_2021_Pockets} establish this result formally, and under conditions that allow for dependence in  
$d_s$.

Recall $w_t=\left(y_t,x_t'\right)'$ and denote the element in position $\iota$ as $\{ w_{\iota,t} \}$.

\textbf{Assumption B} 
\textit{ Let $ \{ w_{\iota,t} \}_{t \geq 1} $  be a strictly stationary sequence of random variables and $\{ v_t(\xi) \}_{t \geq 1} $, 
$\{ v_t(\zeta) \}_{t \geq 1} $  with $\xi, \ \zeta \in \R$, sequences of real numbers.   For each $1 \leq i  \leq j$, set 
$ \mathcal{F}_i^j \left( v_t(\cdot) \right) $ as the $\sigma-$algebra generated by the events $\{ w_{\iota,r} \leq v_t(\cdot)\}$, $i \leq r \leq j$, and, for $1 \leq l \leq t-1$, denote
\begin{equation*}
\alpha_{t,l}(\xi,\zeta) = \max_{1\leq k \leq t-l} \left\{ \left|  P \left(A \cap  B\right)-P(A)P(B) \right|  \ : 
\   A \in  \mathcal{F}_1^k \left( v_t(\xi) \right),  \ B \in  \mathcal{F}_{k+1}^t \left( v_t(\zeta) \right)        \right\},
\end{equation*}
then there exists a sequence $l_t(\xi; \zeta)=o(t)$, as $t \rightarrow \infty$, such that  
$\lim_{t_\rightarrow \infty}\alpha_{t,  l_t}(\xi,\zeta) = 0$.
}

Assumption B is discussed in \citet{Ferreira2002} and is a very mild mixing condition, which relaxes regularity condition $D(u_n)$ in \citet{Leadbetter1983}, page 53, which in turn is already a relaxation of strong mixing. Notice that the stationarity and mixing condition in \citet{Ferreira2002} is referred to the series $d_s$: differently, we formulate this requirement in terms of the observables $w_t$: the validity of this stationarity and mixing conditions for $d_s$ follows from Theorems 3.35 and 2.49 of \citet{white2000asymptotic}, recalling that $R$ is finite so the application of this result is justified.

Assumption B is sufficient to establish Proposition 1 of  \citet{Taylor_etal_2021_Pockets}. We then obtain
\begin{theorem}\label{Thm_HLT}
Under assumption B, as $T \rightarrow \infty$,
\begin{equation*}
\lim P \left(\max_{s=T^*+1,\ldots,E} d_s^2>  \max_{s=1,\ldots,T^*} d_s^2 \right)=\frac{\lambda_2-\lambda_1}{\lambda_2}.
\end{equation*} 
\end{theorem} 
Thus, following \citet{Taylor_etal_2021_Pockets} we suggest to consider $\max_{s=T^*+1,\ldots,E} d_s^2$ and $\max_{s=1,\ldots,T^*} d_s^2 $ and conclude that there is instability if $\max_{s=T^*+1,\ldots,E} d_s^2>  \max_{s=1,\ldots,T^*} d_s^2$. \citet{Taylor_etal_2021_Pockets} refer to this as the MAX procedure, which is a test with size $\frac{\lambda_2-\lambda_1}{\lambda_2}$.

Notice that although the monitoring interval $[\lfloor \lambda_1 T \rfloor+1,\lfloor \lambda_2 T \rfloor]$ is assumed to be proportional to the sample size, we still consider the MAX a diagnostic to detect instability even at a single point in time, or for a very short period, as that could be sufficient to cause $\max_{s=T^*+1,\ldots, E} d_s^2$ to exceed the threshold from $\max_{s=1,\ldots, T^*} d_s^2$.

In a nutshell, consider a sequence $\{\upsilon_s\}$ of independent standard normal distributions, observed at  $s=1,\ldots,T$, with a possible point of instability $\upsilon_{s^*}$, with $E(\upsilon_{s^*})=\mu_s^*$ and ${s^*}>{T^*}$. The MAX procedure compares $\max_{s=T^*+1,\ldots, E} \upsilon_s$ against $\max_{s=1,\ldots, T^*} \upsilon_s$, and it is well-known that the latter has stochastic order $O_p((\ln(T^*))^{1/2})$, see Theorem 1.5.3 in \citet{Leadbetter1983}. Thus, the MAX procedure would have non-trivial power against cases with $\mu_s^*=c (\ln(T^*)^{1/2})$ when $c>0$. Interestingly, this is slightly less than the local power for the test based on \citet{andrews2003end} (which has non-trivial power if $\mu_s^*>0$), but much more than the local power of the Fluctuation test or of the Diebold and Mariano test (as we have seen these have no power in this situation). Thus, we state that the \citet{andrews2003end} test and the MAX procedure may complement the information from the Fluctuation and DM tests. 

\section{Monte Carlo Studies}\label{Sec_MC} 

This section is dedicated to investigating the size and properties of the tests previously described in a complete Monte Carlo study. As a second experiment, we compare the different versions of the Andrews ($S$) test when different choices of $\Sigma$ are considered. 
In particular, we consider the Data Generating Process (DGP) as 
\begin{align}
    & y_t=\beta x_t +\eta_t,  \label{DGP}\\
    & x_t=\rho_x x_{t-1}+\xi_t \text{,} \ \
    \xi_t \sim N.i.d.(0,\sigma^2_x) \text{,} \ |\rho_x|<1, \nonumber\\
    & \eta_t=\rho_\eta \eta_{t-1}+\varepsilon_t \text{,} \ \ \varepsilon_t \sim N.i.d.(0,\sigma^2_\eta) \text{,} \ |\rho_\eta|<1. \nonumber
\end{align}
We assume that we do not observe $x_t$, but
\begin{align*}
    & x_t^{(1)*} = x_t + v^{(1)}_t,  \ \ v^{(1)}_t \sim  N.i.d.(0,\sigma^2_1),\\
    & x_t^{(2)*} = x_t + v^{(2)}_t,  \ \ v^{(2)}_t \sim  N.i.d.(0,\sigma^2_2),
    \end{align*}
where $\xi_t$, $\varepsilon_{t'}$, $v_{t''}^{(1)}$ and $v_{t'''}^{(2)}$ are independent for each $t$, $t'$, $t''$ and $t'''$. 

Let us denote
\begin{equation*}
\widehat{\beta}_t^{(1)} = \frac{\sum_{l=t-R}^{t-1} y_l x_l^{(1)*}}{\sum_{l=t-R}^{t-1} {\left(x_l^{(1)*}\right)}^2}, \qquad   
\widehat{\beta}_t^{(2)} = \frac{\sum_{l=t-R}^{t-1} y_l x_l^{(2)*}}{\sum_{l=t-R}^{t-1} {\left(x_l^{(2)*}\right)}^2},
\end{equation*}
where $R$ is the rolling window size. Then, we consider the forecasting rule as 
\begin{equation*}
\widehat{y}^{(i)}_t=\widehat{\beta}_t^{(i)} \ x_t^{(i)*}, \qquad \text{ for } i = 1,2.    
\end{equation*}
Hence, this could be a situation in which forecaster $i$ is trying to predict an unobserved factor $x_t$ but she/he does it with an error, as it may happen when unobservable factors are estimated by using only a small information set.

Thus, the forecast errors are defined as
\begin{align*}
     & e^{(i)}_t=\beta x_t + \eta_t - \widehat{\beta}_t^{(i)} x_t - \widehat{\beta}_t^{(i)} v_t^{(i)}, \quad \text{ for } i =1,2,
\end{align*}
and notice that, for $i=1, 2$, $\widehat{\beta}_t^{(i)}$ is independent from $v_t^{(i)}$, as the latter is serially independent, and $v_t^{(2)}$ is independent from $x_t$. Then, the equal predictive ability useful for the DM test is 
\begin{align}
     & E\left(e^{(1)}_t\right)^2-E\left(e^{(2)}_t\right)^2 \nonumber \\
&=E \left\{ \left(\beta-\widehat{\beta}_t^{(1)}\right)^2 x_t^2 \right\} - E \left\{ \left(\beta-\widehat{\beta}_t^{(2)}\right)^2 x_t^2 \right\}
+  E  \left(\widehat{\beta}_t^{(1)}\right)^2 \sigma_1^2 - E  \left(\widehat{\beta}_t^{(2)}\right)^2 \sigma_2^2 \nonumber \\
&+ 2 E \left\{ \left({\beta}-\widehat{\beta}_t^{(1)} \right) x_t \eta_t  \right\}-2 E \left\{ \left({\beta}-\widehat{\beta}_t^{(2)}\right) x_t \eta_t  \right\}, \label{eq:Simu1}
    \end{align}
where the elements in Eq.~\eqref{eq:Simu1} are equal to $0$. 

If  $\sigma_1^2=\sigma_2^2$, then $\left\{ v_{t-R}^{(1)},\ldots, v_{t-1}^{(1)} \right\}'$ and $\left\{ v_{t-R}^{(2)},\ldots, v_{t-1}^{(2)}\right\}'$ are identically distributed, and  $\widehat{\beta}_t^{(1)}$, $\widehat{\beta}_t^{(2)}$, as a finite function of identically distributed variables, are also identically distributed, so $E  \left(\widehat{\beta}_t^{(1)}\right)^2 \sigma_1^2 =  E  \left(\widehat{\beta}_t^{(2)}\right)^2 \sigma_2^2$. 
Similarly, it follows that $E \left\{ \left(\beta-\widehat{\beta}_t^{(1)}\right)^2 x_t^2 \right\} = E \left\{ \left(\beta-\widehat{\beta}_t^{(2)}\right)^2 x_t^2 \right\}$ and
$E \left\{ \left({\beta}-\widehat{\beta}_t^{(1)}\right) x_t \eta_t  \right\}= E \left\{ \left({\beta}-\widehat{\beta}_t^{(2)}\right) x_t \eta_t  \right\}$. It is straightforward to verify that $d_t$ is strictly stationary, strong mixing, and ergodic.

To evaluate the power of these tests it is sufficient to modify the value of $\sigma_2^2$ for one or more points $t$.
In our baseline design, we set
\begin{align} \label{baseline}
\beta &= 1, \quad \rho_x = 0.75, \quad  \sigma_x^2 = 1, \\
\rho_\eta &= 0.5, \quad  \sigma_\eta^2 = 0.1, \quad  \sigma_1^2 = 0.1, \quad  \sigma_2^2 = 0.1 \cdot \delta_s, 
\end{align}
where $\delta_s=1$ yields $E(d_s)= 0$. 

We simulate the DGP in Eq.~\eqref{DGP} with parameters as in \eqref{baseline}, with sample size $T$ equal to $80$ and rolling window size $R$ equal to $20$ as in our real data application. Moreover, we set $\kappa=0.3$ for the Fluctuation test, and $T^*=76$ and $E=80$ for the $MAX$ diagnostic. For the DM and fluctuation tests, we estimate the long-run variance using the Bartlett kernel with bandwidth $M=\lfloor T^{2/9} \rfloor$, also we use fixed smoothing critical value for the DM test.

Subsequently, we compute the $DM$, fluctuation ($Fl$), and $S$ statistics and perform the corresponding tests, alongside the $MAX$ procedure.  
In detail, we set $\delta_s=1$ to study the size, and three different specifications for the power: 
\begin{itemize}
    \item[(i)] $\delta_s = \delta \neq 1$ for all $s$, which results in $E(d_s)\neq 0$ for all $s$. It is the usual assumption underpinning the motivation for the DM test.
    \item[(ii)] $\delta_s = \delta \neq 1$ for $s>\lfloor \tau T\rfloor$ and 1 otherwise (we used $\lfloor \tau T\rfloor=T-20$), which results in $E(d_s)\neq 0$ for $s>\lfloor \tau T\rfloor$. It is an example of those local deviations from the $E(d_s)=0$ hypothesis that motivated the introduction of the Fluctuation test.
    \item[(iii)] $\delta_T = \delta \neq 1$ only when $s=T$ and 0 otherwise, which is considered the short deviation. This situation may describe well what happened when the COVID-19 shock hit the economy, which we may interpret as saying that when $\delta>1$ the factor estimate $x_T^{(2)*}$ is less precise than $x_T^{(1)*}$.  
\end{itemize}
For each experiment, we run $10.000$ replications. Notice that $\delta_s = \delta \neq 1$  also affects the variance of $d_s$: this is particularly relevant for the $S$ test and $MAX$ procedure.
Regarding the theoretical size, we set $5\%$. The summary of the results is provided in Table~\ref{tab:MC_all}.

\begin{table}[b!] \centering
    \captionsetup{font=small}
    \caption{
    Global and local Equal predictive ability tests for different sizes and power. We report in columns the DM, fluctuation, and $S$ test and the \text{MAX} procedure when $T = 80$ and $R = 20$.}
    \label{tab:MC_all}
    \small
        \begin{tabular*}{\linewidth}{@{\extracolsep{\fill}} cccccc}

\toprule
Size/Power                       & $\delta$ & DM & $Fl$ & S & MAX \\[+2pt]
\toprule
$\delta_s = \delta$ for all $s$  & 1 & 0.053 & 0.047 & 0.046 & 0.051  \\
\midrule
$\delta_s = \delta$ for all $s$  &0.1   & 0.919 & 0.654 & 0.051 & 0.054 \\
                                 &2     & 0.925 & 0.658 & 0.049 & 0.051 \\
                                 &4     & 1.000 & 0.986 & 0.048 & 0.056  \\
\midrule
$\delta_s = \delta$ for $s>T-20$ & 0.1   & 0.091 & 0.081 & 0.029 & 0.029  \\
                                 &2     & 0.160 & 0.249 & 0.106 & 0.150  \\
                                 & 4     & 0.547 & 0.884 & 0.150 & 0.150  \\
\midrule
$\delta_s = \delta$ for $s=T$    & 0.1   & 0.053 & 0.048 & 0.024 & 0.044 \\ 
                                 &  2     & 0.052 & 0.044 & 0.167 & 0.114 \\
                                 & 4     & 0.047 & 0.034 & 0.426 & 0.335 \\
                                 & 8     & 0.034 & 0.021 & 0.672 & 0.609  \\
\bottomrule
    
        \end{tabular*} \\ \smallskip
        \scriptsize
        Note: the table exhibits the empirical size and power of the equal predictive ability tests. The theoretical size is set at 5\% for all the tests.
\end{table}

All the tests are well-sized under the null hypothesis. The local power depends on the nature of the violation of the $E(d_s)=0$  hypothesis. 
The DM test (third column) has the best power performance when the deviation $E(d_s) \neq 0$ holds for all $s$, whereas the fluctuation test (forth column) has the best power performance in the case of a local but extended deviation. The $S$ test and $MAX$ diagnostic (fifth and sixth column) have no power when the deviation from the null hypothesis occurs at every point, and not much power (compared to the fluctuation test) in case of local but extended deviations. 
On the other hand, the $S$ test and $MAX$ diagnostic have the best power in the case of a very short deviation: interestingly, the $S$ has marginally more power compared to the $MAX$, reflecting the fact the $S$ test assumes exact knowledge of the location of the shift, whereas the $MAX$ procedure is more agnostic about this piece of information. On the other hand, the DM and fluctuation tests have no effective power in this situation of very short deviation.

Finally, we notice that, for the case of global and local deviations, the power of the DM and fluctuation tests increase with the distance from the null hypothesis $\delta=1$; the same holds for the $S$ test and $MAX$ diagnostic in the presence of very short deviation, but only when $\delta>1$: we conjecture that the lack of power for the  $S$ test and $MAX$ procedure when $\delta<1$ depends on the reduction in the variance in $d_t$. 


In the second experiment, we compare three versions of the  $S$ test when more than one period is considered: in the first case (Column I), we do not apply a correction for the dependence in the test statistic, corresponding to using the identity matrix; in the second case (Column $\widetilde{\Sigma}$), we use $\widetilde{\Sigma}$ as recommended in \cite{andrews2003end}, and in the third case (Column $\widehat{\Sigma}$), we employ $\widehat{\Sigma}$, where $\Sigma$ is estimated only using the pre-change sample. 

The outcome of the experiment is provided in Table~\ref{tab:MC_Sigmas}: as \cite{andrews2003end} anticipated using $\widehat{\Sigma}$ yields slightly better empirical size, since a few more observations are used; however, this size correction results in much less power when the dimension of the instability is large. On the other hand, perhaps because the dependence in the Monte Carlo exercise is not very large, we find that the power gain from using the estimate of $\Sigma$ instead of the Identity matrix is negligible.   

\begin{table}[h]
  \centering
  \caption{Size and Power of the $S$ tests for various values of $\delta$ and estimates of $\Sigma$ (equal to $I$, $\widetilde{\Sigma}$ and $\widehat{\Sigma}$). We consider $T = 80$, a period of instability equal to $3$, and a theoretical size equal to $5\%$.}
    \begin{tabular}{cccc}
\toprule
$\delta$ & I  & $\widetilde{\Sigma}$ & $\widehat{\Sigma}$ \\[+2pt]
\toprule
    1     & 0.058 & 0.053 & 0.059 \\
    2     & 0.273 & 0.250 & 0.273 \\
    4     & 0.715 & 0.665 & 0.714 \\
    8     & 0.944 & 0.783 & 0.944 \\
    16    & 0.992 & 0.717 & 0.992 \\
    32    & 0.999 & 0.731 & 0.999 \\
    64    & 1.000 & 0.794 & 1.000 \\
    128   & 1.000 & 0.854 & 1.000 \\
    \bottomrule
    \end{tabular}
  \label{tab:MC_Sigmas}%
\end{table}%

\section{Evaluating the nowcast of US nominal GDP growth}\label{Sec_Application}

In this section, we illustrate the results obtained from the Monte Carlo exercise with an empirical example dedicated to evaluating the nowcast of US nominal GDP growth by using the GDP nowcast from the Survey of Professional Forecasters (SPF) over the period 2000:Q1 to 2020:Q3. 
In particular, we consider for the SPF the nowcast of nominal GDP, denoted $\widehat{y}_t$, that is made when only information on nominal GDP of the previous quarter is available, ${y}_{t-1}$. As the survey covers many individuals, we use as $\widehat{y}_t$ the median of the responses for each point in time. 

The nowcast of the quarterly growth rate is then 
$\frac{\widehat{y}_t-{y}_{t-1}}{{y}_{t-1}}$ and the error associated with the SPF nowcast is denoted as 
 $e_t^{(1)}=\frac{{y}_{t}-\widehat{y}_t}{{y}_{t-1}}$. As a benchmark, we consider nowcasting nominal GDP growth as 0, which corresponds to nowcasting the GDP as the last available observation, $\widetilde{y}_t={y}_{t-1}$, so the error associated with the benchmark nowcast, $e_t^{(2)}$, is therefore $e_t^{(2)}=\frac{{y}_{t}-\widetilde{y}_t}{{y}_{t-1}}$.   
 Using the quadratic loss function, the loss differential is then $d_t=e_t^{(1)2}-e_t^{(2)2}$. 

Clearly, the benchmark in this exercise is not a very effective nowcast, for example, it even neglects the long-run growth in nominal GDP due both to economic growth and inflation. However, it is a convenient one, in the sense that we expect that $E(d_t)<0$, and therefore the null hypothesis should be rejected. This is then a fitting benchmark to check the predictions from Theorem~\ref{DM}.

\begin{figure}[h!]
    \centering
    \caption{US nominal GDP growth (black dotted), GDP nowcast from the SPF (blue line, denoted as $\widehat{y}_t$), and naive nowcast (red line, $\widetilde{y}_t$) over the period 2000:Q1 to 2020:Q3.} 
    \includegraphics[trim= 0mm 0mm 0mm 0mm,clip, width= 12.0cm]
  {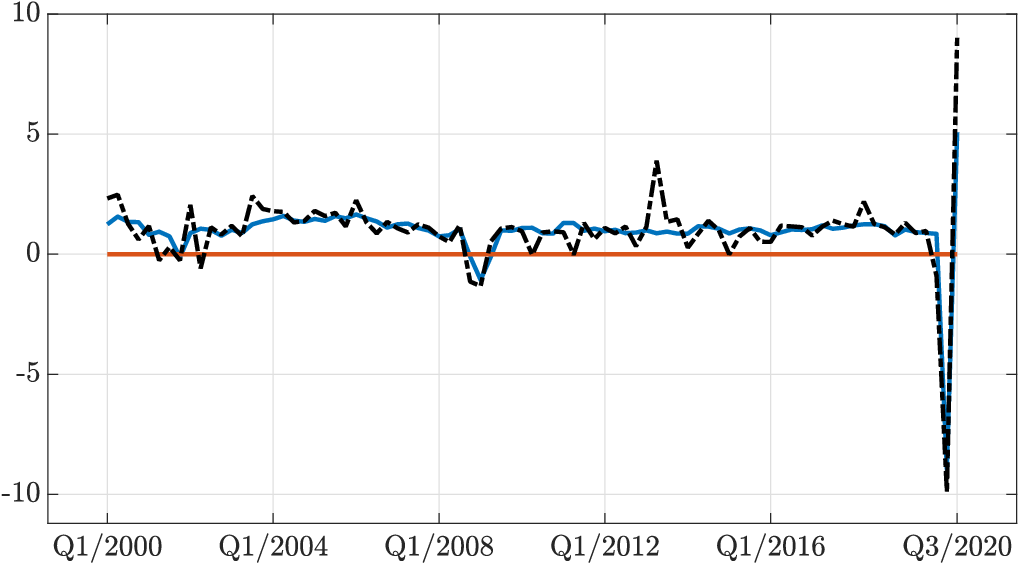}
    \label{fig:SPF-Observed}
\end{figure}

\begin{figure}[h!]
    \centering
    \caption{Errors associated to the SPF nowcast ($e_t^{(1)}$, blue line) and the naive/benchmark model ($e_t^{(2)}$, red line) over the period 2000:Q1 to 2020:Q3.} 
   \includegraphics[trim= 0mm 0mm 0mm 0mm,clip, width= 12.0cm]{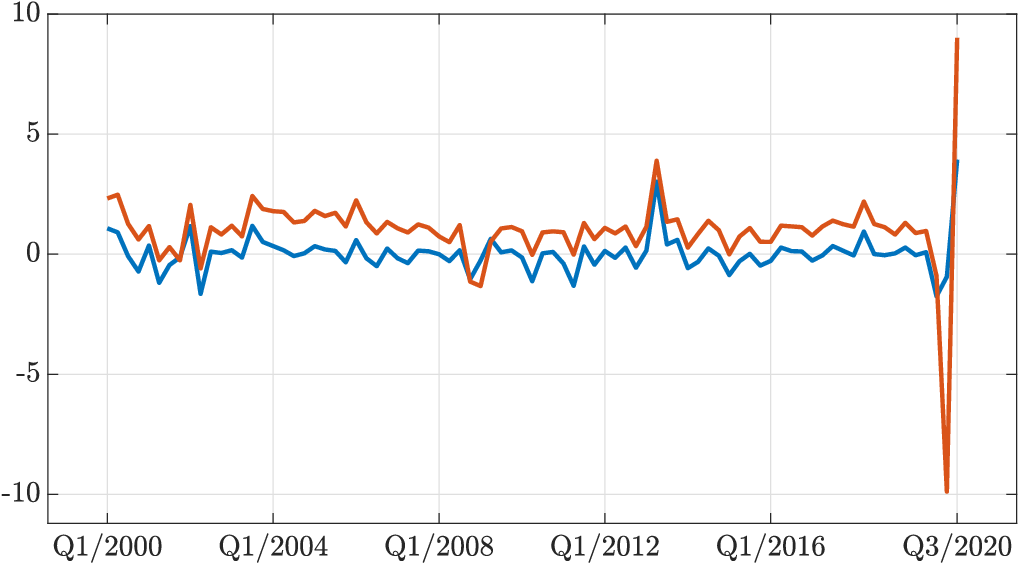}
    \label{fig:SPF-errors}
\end{figure}

Figure~\ref{fig:SPF-Observed} provides the plot of the GDP growth, along with the SPF nowcast and the naive benchmark, while Figure~\ref{fig:SPF-errors} shows the errors.

The SPF nowcast is always close to the target, except for an unanticipated and temporary surge of the growth rate in 2013; as anticipated, the naive benchmark does not take into account long-run real economic growth and inflation and the errors associated with it are then always positive, except for the 2008-2009 recession. Crucially, the SPF tracked well even the 2020 shock, whereas the naive errors are much higher for that period (in absolute value). 

As a result (see Table~\ref{tab:Evaluation}), the performance in root mean square error terms (RMSE) of the naive benchmark worsens compared to the SPF nowcast, the ratio of the two RMSEs passing from 0.22 over the period up to 2019, to 0.16 when the three quarters in 2020 are added, even though the RMSE increased for both sources. 
We then turn to the DM and the Fl tests, again evaluated with the RMSE loss function: in both cases, we estimated the long-run variance using the Bartlett kernel and bandwidth $\lfloor T^{2/9}\rfloor$. For the Fluctuation test, we set $k$ so that the $k/T$ ratio, $\kappa$, is $0.3$, so the critical value for $FL_\kappa$ is $3.012$.  

A summary of the results is available in Table~\ref{tab:Evaluation}. A negative entry for the DM statistic means that the average RMSE for the SPF nowcast is lower than the average RMSE of the benchmark; $Fl_l$ and $FL_u$ are the minimum and maximum of the statistic $Fl_{s,k}$, respectively, where again a negative entry refers to the situation in which the RMSE of the benchmark exceeds the RMSE of the SPF nowcast, so the latter is more precise. In this example, both the DM and the Fluctuation tests suggest that the SPF nowcast is more precise than the benchmark, and significantly so if the sample is limited to up to 2019. As we anticipated the SPF does not seem to outperform the naive benchmark using the DM and the Fl tests when the observations for 2020 are included in the sample, although the RMSE ratio is even more favorable. 

\begin{table}[h!]
      \centering    
        \caption{Nowcast evaluation based on the SPF and the naive benchmark through the RMSE, the DM, and Fluctuation test over the period 2000:Q1 to 2019:Q4 and 2000:Q1 to 2020:Q3.}
        \label{tab:Evaluation}
    \begin{tabular*}{\linewidth}{@{\extracolsep{\fill}} lcccccc}
        \midrule
        Period & $\text{RMSE}_{\text{SPF}}$ & $\text{RMSE}_{\text{Naive}}$ & Ratio & $DM$ & $Fl_{l}$ & $Fl_{u}$ \\ [1pt]
        \midrule
            Q1:2000 - Q4:2019 & 0.37   & 1.66   & 0.22  & -7.27 & -5.83 & -2.04 \\
            Q1:2000 - Q3:2020 & 0.59   & 3.77   & 0.16  & -1.92 & -2.49 & -0.21 \\
        \midrule        
    \end{tabular*} \\
    \scriptsize
    Note: Columns $\text{RMSE}_{\text{SPF}}$ and $\text{RMSE}_{\text{Naive}}$ are the average RMSE for the SPF and Naive benchmark, respectively. Column Ratio refers to the ratio $\text{RMSE}_{\text{SPF}}/\text{RMSE}_{\text{Naive}}$. Columns $DM$, $Fl_l$, and $Fl_u$ are the DM and lower and upper Fl test statistics. 
    The $5\%$ critical values for two-sided tests are 2.032 (fixed smoothing) and 3.012, respectively.
\end{table}

In conclusion, we analyse the performance of the two nowcasting models using the $S$ Andrews test and the $MAX$ diagnostic over the COVID-19 period (2020:Q1 - 2020:Q3). As the location of the COVID-19 recession and recovery can be treated as a shock at a known date, the application of the $S$ test seems appropriate, and in our case we focus on just three observations.

Results for the two diagnostics are in Table \ref{tab:Andrews2003Max}: these both suggest an increase of the forecasting differential in the second period, but for the $S$ statistic weighing the errors with the restricted residuals we fail to reject the null, consistently with the finding in the Monte Carlo experience that this has less power, especially in presence of large shocks.  Let us also add that the last column $q_{\text{MAX}}$ is the critical value which coincides with the MAX procedure over the pre-covid period (2000:Q1 - 2019:Q4).

\begin{table}[h!]
    \centering
    \caption{\cite{andrews2003end} $S$ test evaluated for three different values of $\Sigma$ and MAX procedure over the period 2020:Q1 to 2020:Q3 (3 observations). 
    }
    \label{tab:Andrews2003Max}
    \begin{tabular*}{\linewidth}{@{\extracolsep{\fill}} cccccccc}
       \midrule
        $S(I)$  &  $q_{S(I)}$ & 
        $S(\widetilde{\Sigma})$  &  $q_{S(\widetilde{\Sigma})}$& 
        $S(\widehat{\Sigma})$  &  $q_{S(\widehat{\Sigma})}$&
        $MAX$ & $q_{MAX}$ \\
        \midrule
          7576   &10.9  &  0.21&    1.92& 3060&3.6  &  $96.84^2$  &$6.03^2$   \\
        \midrule 
    \end{tabular*} \\
        \scriptsize
        Note: Columns $S(I)$, $S(\widetilde{\Sigma})$, and 
        $S(\widehat{\Sigma})$,
        denote the $S$ test statistics when the Identity matrix, the restricted residuals, and the unrestricted residuals are used to weight the errors, respectively. Columns $q_{S(I)}$, $q_{S(\widetilde{\Sigma})}$, and $q_{S(\widetilde{\Sigma})}$, 
         are the respective critical values. The theoretical size is 5\%. Column MAX denotes the maximum procedure over the period 2020:Q1 to 2020:Q3, while column $q_{\text{MAX}}$ denotes the MAX over the period 2000:Q1 to 2019:Q4. The false positive rate of the procedure is 3.6\%.
\end{table}

\section{Conclusions}\label{Sec_Conclusions} 
 COVID-19 was an exceptionally challenging event for forecasting and evaluation since it was a moment of extreme instability spanning over a very short period.
In particular, tests like the \cite{dieboldComparingPredictiveAccuracy2002} test for equal forecasting ability or the \cite{giacomini2010forecast} test have less power, if the time span of the instability is very short, even when the local discontinuity is very large relative to the sample size. 

In this paper, we show that in these situations and subsequent recovery, using non-parametric diagnostics (such as the $S$ test or the MAX procedure) for local breaks or extreme values leads to better conclusions.
We illustrate these results in two Monte Carlo exercises, and we provide evidence of the importance of selecting the correct test in a nowcasting exercise for the nominal US GDP, where we compare the SPF and a naive benchmark.
Given these results, we recommend that the forecaster should not pool the sample, but exclude the short periods of high local instability from the evaluation exercise.

\pagebreak
\bibliographystyle{chicago}
\bibliography{reference}

\newpage
\appendix
\renewcommand{\theproposition}{\thesection.\arabic{proposition}}
\renewcommand{\theequation}{\thesection.\arabic{equation}}
\setcounter{equation}{0}

\section{Appendix} \label{sec:AppA}

     \textbf{Proof of Theorem~\ref{DM}} \\
Let us denote $\nu_s=\delta_2 T^a I_s(\tau)$, then we rewrite $c_l$ as 
\begin{equation*}
    c_l=c_l(uu)+c_l(u \nu)+c_l(\nu u)+c_l(\nu \nu),
\end{equation*}
where
\begin{align*}
    & c_l(uu)=\frac{1}{T} \sum_{s=l+1}^T {(u_s-\overline{u})(u_{s-l}-\overline{u})}, \\
    &c_l(u \nu)=\frac{1}{T} \sum_{s=l+1}^T {(u_s-\overline{u})(\nu_{s-l}-\overline{\nu})}, \\
    & c_l(\nu u)=\frac{1}{T} \sum_{s=l+1}^T {(\nu_s-\overline{\nu})(u_{s-l}-\overline{u})}, \\
    &c_l(\nu \nu)=\frac{1}{T} \sum_{s=l+1}^T {(\nu_s-\overline{\nu})(\nu_{s-l}-\overline{\nu})}.
\end{align*}
So 
\begin{equation*}
 \widehat{\sigma}^2_T= \widehat{\sigma}^2_T(uu)+ \widehat{\sigma}^2_T(\nu u)+  \widehat{\sigma}^2_T(u \nu)+
  \widehat{\sigma}^2_T(\nu \nu),
\end{equation*}
where 
\begin{equation*}
\widehat{\sigma}^2_T(uu) =
 c_0(uu)+2 \sum_{l=1}^M {\frac{M-l}{M} c_l(uu)},
\end{equation*}
and $\widehat{\sigma}^2_T(\nu u)$,   $\widehat{\sigma}^2_T(u \nu)$,
  $\widehat{\sigma}^2_T(\nu \nu)$ 
  are defined in the same manner. \\
  Under Assumptions GW.1 - GW.3, $\widehat{\sigma}_T(uu) - {\sigma}_T \rightarrow_p 0$ as in Theorem 4 of \cite{GiacominiWhite2006}.  \\
  For the contribution to $\widehat{\sigma}^2_T(\nu \nu)$, first notice that 
  \begin{equation*}
      \overline{\nu}=\frac{1}{T}\sum_{s=1}^T \nu_s=\frac{1}{T} \delta_2 T^a.
  \end{equation*}
  Thus,
  \begin{align*}
    & c_0(\nu \nu)=\frac{1}{T}\sum_{s=1}^T \left(\nu_s-\overline{\nu}\right)^2=\frac{1}{T}\sum_{s=1}^T \nu_s^2 - \overline{\nu}^2  \\ 
    & = \frac{1}{T}\left(\delta_2 T^a\right)^2 - \left(\frac{1}{T} \delta_2 T^a\right)^2=\delta_2^2 T^{2a-1}-\delta_2^2 T^{2a-2}=\delta_2^2 T^{2a-1}+o\left(T^{2a-1}\right).
  \end{align*}
Looking at $c_l(\nu \nu)$, for $\tau \in (0,1)$, and for $T$ large enough, then $l<\lfloor \tau T \rfloor < T-l$, and
\begin{align*}
    & \frac{1}{T} \sum_{s=l+1}^T \nu_s \nu_{s-l}=0, \\
     & - \frac{1}{T} \sum_{s=l+1}^T \nu_s \overline{\nu}=  - \frac{1}{T} \left(\delta_2 T^a\right)\left(\frac{1}{T} \delta_2 T^a\right)=-\delta_2^2 T^{2a-2}, \\
    & - \frac{1}{T} \sum_{s=l+1}^T \nu_{s-l} \overline{\nu}=  - \frac{1}{T} \left(\delta_2 T^a\right)\left(\frac{1}{T} \delta_2 T^a\right)=-\delta_2^2 T^{2a-2},    \\
    & \frac{1}{T} \sum_{s=l+1}^T  \overline{\nu}^2=\frac{1}{T}(T-l)\left(\frac{1}{T} \delta_2 T^a\right)^2=
    \frac{T-l}{T} \delta_2^2 T^{2a-2},
\end{align*}
we obtain
 \begin{equation*}
     c_l(\nu \nu)=-\delta_2^2 T^{2a-2} \
- \frac{l}{T} \delta_2^2 T^{2a-2}=
-\delta_2^2 T^{2a-2}+o\left(T^{2a-2}\right)
 \end{equation*} 
 so  
  \begin{equation*}
    \widehat{\sigma}^2_T(\nu \nu)=
    \delta_2^2 T^{2a-1} - M \delta_2^2 T^{2a-2}+o\left(T^{2a-1}+M T^{2a-2}\right)
    =\delta_2^2 T^{2a-1}+o\left(T^{2a-1}\right).
 \end{equation*}
 The approximation $\widehat{\sigma}^2_T(\nu \nu) = \delta_2^2 T^{2a-1}+o\left(T^{2a-1}\right)$ when $\tau=0$ or $\tau=1$ can be established in the same way, using $\frac{1}{T} \sum_{s=l+1}^T \nu_{s-l} \overline{\nu}=0$ when $\tau=1$, and 
$\frac{1}{T} \sum_{s=l+1}^T \nu_s \overline{\nu}=0$ when $\tau=0$.

\noindent \textbf{Case $a<1/2$:} \\
We obtain $\widehat{\sigma}_T(\nu \nu)^2 = o(1)$, and, by the Cauchy-Schwarz inequality, $\widehat{\sigma}^2_T(u \nu) = o_p(1)$, and $\widehat{\sigma}^2_T(\nu u) = o_p(1)$, therefore 
 $\widehat{\sigma}_T^2-\sigma_T^2\rightarrow_p 0$. As for the contribution of $\nu_s$ to the numerator of the DM statistic, 
 \begin{equation*}
   \frac{\sqrt{T}}{T} \sum_{s=1}^T \nu_s = T^{-1/2} \delta_2 T^a =o(1)  
 \end{equation*}
 so 
  \begin{equation*}
   t_{DM} 
   \rightarrow_d Z+\frac{\delta_1}{\sigma}.
 \end{equation*}

\noindent \textbf{Case $a>1/2$:} \\
We obtain $T^{1-2a} \widehat{\sigma}_T(\nu \nu)^2\rightarrow \delta_2^2$, and $T^{1-2a} \widehat{\sigma}_T(u u)^2\rightarrow_p 0$, so,  by the Cauchy-Schwarz inequality, $\widehat{\sigma}_T(u \nu)^2 = o_p(1)$, and $\widehat{\sigma}_T(\nu u)^2 = o_p(1)$, therefore 
 $T^{1-2a} \widehat{\sigma}_T^2 \rightarrow_p \delta_2^2$. As for the contribution of $\nu_s$ to the numerator of the DM statistic, 
 \begin{align*}
  & \frac{T^{1/2-a} \sqrt{T}}{T} \sum_{s=1}^T \nu_s \rightarrow \delta_2,  \\
  & \frac{T^{1/2-a} \sqrt{T}}{T} \sum_{s=1}^T \left(u_s+\delta_1 T^{-1/2}\right) \rightarrow_p 0,
 \end{align*}
 so 
  \begin{equation*}
   t_{DM} 
   \rightarrow_p \frac{\delta_2}{|\delta_2|}.
 \end{equation*}
\qed 
\bigskip \\
 \textbf{Proof of Theorem \ref{Thm_Andrews}} \\
 To prove this theorem, we need to show that the assumptions A.1-A.4 correspond to similar assumptions in \cite{andrews2003end}.
 
 As $\widehat{y}^{(i)}_{t}\left(\widehat{\delta}^{(i)}_{t-h,R_i}\right)$ is a function of $\{ w_{t-h}, w_{t-h-1}, \hdots, w_{t-h-R_i+1} \}$ in view of Theorem 3.35 of \citet{white2000asymptotic}, then $\widehat{y}_{t}^{(i)}\left(\widehat{\delta}^{(i)}_{t-h,R_i}\right)$ are also stationary and ergodic. Similarly,  ${e}_{t}^{(i)}\left(\widehat{\delta}^{(i)}_{t-h,R_i}\right)$,  $L\left({e}_{t}^{(i)}\left(\widehat{\delta}^{(i)}_{t-h,R_i}\right)\right)$, and $d_{t}\left(\left(\widehat{\delta}^{(1)}_{t-h,R_1}\right),\left(\widehat{\delta}^{(2)}_{t-h,R_2}\right)\right)$ are also stationary and ergodic.   
\\
Assumptions A.1 and A.2 are sufficient to establish Assumption 1 in \citet{andrews2003end}, and Assumptions A.3 and A.4 correspond to similar assumptions in sufficient condition LS \citet{andrews2003end}, where, however, our situation is simpler because in our restricted model we only have a regression on a constant.
\qed
\end{document}